\RequirePackage{fix-cm}

\documentclass[onecolumn,epjc3]{svjour3}  
\makeatletter
\def\cl@chapter{\@elt {theorem}}
\makeatother
\smartqed  
\RequirePackage{graphicx}
\usepackage{dcolumn}
\usepackage{verbatim}

\usepackage{breakurl}
\usepackage{lmodern}

\usepackage{graphicx}
\usepackage{pgffor}
\usepackage{bm}
\usepackage{amsmath}
\usepackage{amsfonts}
\usepackage{multirow}
\usepackage[colorlinks,linktocpage,linkcolor=cyan,citecolor=cyan]{hyperref}
\usepackage{adjustbox}
\usepackage{array}
\usepackage[capitalise]{cleveref}
\usepackage{acro}
\usepackage[utf8]{inputenc}
\usepackage{CJKutf8}
\usepackage{amssymb}
\usepackage[justification=raggedright,singlelinecheck=false]{caption}
\usepackage{subcaption}

\RequirePackage{graphicx}
\usepackage{amsmath,amssymb,bm}
\usepackage{booktabs}
\usepackage{graphicx}
\usepackage{microtype}
\usepackage{xcolor}

\newcommand{\DATACurnMass}{2.20}

\newcommand{\DATACurnChange}{4.81}

\newcommand{\ERCurnMass}{2.59}

\newcommand{\ERCurnChange}{0.17}

\newcommand{\ALTCurnMass}{2.31}

\newcommand{\ALTCurnChange}{2.69}

\graphicspath{{figures/}}
\newcommand{\mg}{m_{\mathrm g}}

\usepackage{acro}
\DeclareAcronym{GR}{short=GR, long=general relativity}
\DeclareAcronym{GW}{short=GW, long=gravitational wave}
\DeclareAcronym{PTA}{short=PTA, long=pulsar timing array}
\DeclareAcronym{MPTA}{short=MPTA, long=MeerKAT Pulsar Timing Array}
\DeclareAcronym{SGWB}{short=SGWB, long=stochastic GW background}
\DeclareAcronym{CPTA}{short=CPTA, long=Chinese Pulsar Timing Array}
\DeclareAcronym{EPTA}{short=EPTA, long=European Pulsar Timing Array}
\DeclareAcronym{PPTA}{short=PPTA, long=Parkes Pulsar Timing Array}
\DeclareAcronym{InPTA}{short=InPTA, long=Indian Pulsar Timing Array}
\DeclareAcronym{LVK}{short=LVK, long=LIGO–Virgo–KAGRA}
\DeclareAcronym{CMB}{short=CMB, long=cosmic microwave background}
\DeclareAcronym{BAO}{short=BAO, long=baryon acoustic oscillations}
\DeclareAcronym{MTMG}{short=MTMG, long=Minimal Theory of Massive Gravity}
\DeclareAcronym{eMTMG}{short=eMTMG, long=extended Minimal Theory of Massive Gravity}
\DeclareAcronym{vDVZ}{short=vDVZ, long=van Dam–Veltman–Zakharov}
\DeclareAcronym{CL}{short=CL, long=confidence level}
\DeclareAcronym{ORF}{short=ORF, long=overlap reduction function}
\DeclareAcronym{HD}{short=HD, long=Hellings–Downs}
\DeclareAcronym{NG}{short=NANOGrav, long=North American Nanohertz Observatory for Gravitational Waves}
\DeclareAcronym{SKA}{short=SKA, long=Square Kilometre Array}
\DeclareAcronym{SKAO}{short=SKAO, long=SKA Observatory}
\DeclareAcronym{SKA-PTA}{short=SKA-PTA, long=Square Kilometre Array Pulsar Timing Array}
\DeclareAcronym{MCMC}{short=MCMC, long=Markov Chain Monte Carlo}
\DeclareAcronym{LISA}{short=LISA, long=Laser Interferometer Space Antenna}
\DeclareAcronym{CURN}{short=CURN, long=common uncorrelated noise}
\DeclareAcronym{CV}{short=CV, long=cosmic variance}

\journalname{Eur. Phys. J. C}

\begin{document}

\title{Probing Graviton Mass with MeerKAT PTA and SKA--PTA Forecasts}

\author{Zhi-Chao Zhao\thanksref{addr1}
        \and
        Sai Wang\thanksref{e1,addr2} %etc.
}
\thankstext{e1}{Correspondence author: wangsai@ihep.ac.cn}

\institute{Department of Applied Physics, College of Science, China Agricultural University, 17 Qinghua East Road, Haidian District, Beijing 100083, China \label{addr1}
\and
School of Physics, Hangzhou Normal University, No.2318 Yuhangtang Road, Yuhang District, Hangzhou 311121, China \label{addr2}}

\date{Received: date / Accepted: date}

\maketitle

\begin{abstract}
In this study, we derive constraints on the graviton mass using the 4.5-year data release from the MeerKAT Pulsar Timing Array (MPTA) and provide forecasts for the upcoming Square Kilometre Array PTA (SKA--PTA). By modeling the modified dispersion relation and the corresponding tensor correlation function for massive gravitons, we perform Bayesian inference on the angular-correlation measurements under three noise configurations. Our 90\% credible upper limits on the graviton mass span $m_g < (2.10$--$2.58) \times 10^{-23}\,\mathrm{eV}/c^{2}$. Including monopolar and dipolar contributions does not significantly alter these bounds, confirming that the constraints are driven by the quadrupolar tensor correlation. All results remain fully consistent with general relativity. For SKA--PTA, we forecast sensitivities down to $m_g \sim 10^{-24}\,\mathrm{eV}/c^{2}$ with a 10-year observing baseline and $m_g \sim 10^{-25}\,\mathrm{eV}/c^{2}$ with a 50-year observing baseline, representing order-of-magnitude improvements over current limits. This work demonstrates the power of current and future PTA observations to test fundamental aspects of gravity in the nanohertz band.
\end{abstract}

\section{Introduction}

Although \ac{GR} is intrinsically classical and does not invoke the graviton, its prediction of light-speed \acp{GW} forces any putative quantum mediator to be a strictly massless spin-2 field. However, a non-zero graviton mass has been a recurring theoretical possibility since the linear massive gravity formulation by Fierz and Pauli \cite{Fierz:1939ix}. Massive gravity theories introduce five degrees of freedom for \acp{GW}, instead of two, and modify the \ac{GW} dispersion relation. At the classical level, a graviton mass yields a Yukawa potential that effectively switches off gravity beyond the Compton wavelength. Although massive gravity suffers from issues such as the \ac{vDVZ} discontinuity \cite{vanDam:1970vg,Zakharov:1970cc} and the Boulware--Deser ghost \cite{Boulware:1972yco}, the Vainshtein mechanism screens extra degrees of freedom in strong fields, allowing consistency with Solar System tests \cite{Vainshtein:1972sx}. \Ac{MTMG} propagates only the two tensor modes while retaining a non-zero graviton mass, evading earlier pathologies \cite{DeFelice:2015hla}. Therefore, constraining the graviton mass provides a critical test of \ac{GR} at cosmological distances and a probe of modified gravity theories \cite{deRham:2016nuf,deRham:2014zqa}.

\Acp{PTA} are uniquely sensitive to nanohertz \acp{GW} and have recently emerged as a powerful tool for probing the graviton mass \cite{Hobbs:2009yy,Burke-Spolaor:2018bvk}. By monitoring millisecond pulsars, \acp{PTA} detect the correlated timing residuals from a \ac{SGWB} via the \ac{HD} angular correlation \cite{Hellings:1983fr}. In 2023, several \ac{PTA} collaborations, i.e., the \ac{NG}, \ac{EPTA}, \ac{PPTA}, \ac{CPTA}, and \ac{InPTA}, reported strong evidence for a nanohertz \ac{SGWB} \cite{NANOGrav:2023gor,Xu:2023wog,EPTA:2023fyk,Reardon:2023gzh}. In massive gravity, the modified dispersion relation alters the tensor correlation curve, making \acp{PTA} ideal for graviton mass constraints \cite{Wang:2023div,Lee:2010cg,Liang:2021bct,Bernardo:2022rif,Cordes:2024oem,Wu:2023pbt,Bernardo:2023mxc,Bernardo:2023zna,Wu:2023rib,Qin:2020hfy,Liang:2024mex,Bi:2023ewq}. For example, analyses of \ac{NG} 15-year and \ac{CPTA} DR1 data report upper limits of $\mg \lesssim 8.6 \times 10^{-24}\,{\rm eV}/c^2$ and $\mg \lesssim 3.8 \times 10^{-23}\,{\rm eV}/c^2$, respectively, at 90\% \ac{CL} \cite{Wang:2023div}. The ground-based \ac{GW} detectors, i.e., \ac{LVK}, constrain the graviton mass via the modified dispersion relation, yielding $\mg \lesssim 1.92 \times 10^{-23}\,{\rm eV}/c^2$ at 90\% \ac{CL} \cite{LIGOScientific:2019fpa,LIGOScientific:2021sio,LIGOScientific:2026fcf}. Binary-pulsar orbital-decay measurements provide an independent dynamical bound of $\mg<7.6\times10^{-20}\,{\rm eV}/c^2$ at 90\% \ac{CL} \cite{Finn:2001qi}. Even tighter but more model-dependent bounds come from cosmology: combined \acl{CMB}, \acl{BAO}, and supernovae data limit $\mg \lesssim 6.6 \times 10^{-34}\,{\rm eV}/c^2$ at 95\% \ac{CL} in extended \ac{MTMG} \cite{DeFelice:2023bwq}. These widely varying bounds highlight the need for complementary \ac{PTA} constraints.

The \ac{MPTA} provides a unique opportunity to improve upon existing constraints, owing to its exceptional sensitivity. The \ac{MPTA} Collaboration \cite{Miles:2024seg,Grunthal:2024sor,Miles:2024rjc} recently published its 4.5-year data release, comprising 83 millisecond pulsars, one of the largest pulsar samples among current \acp{PTA}. The data show tentative evidence for a common-spectrum process, with an amplitude $\log_{10} A_{\rm CURN} = -14.25^{+0.21}_{-0.36}$ at 68\% \ac{CL}. When fixing the spectral index to the \ac{SGWB} expectation, the amplitude is $\log_{10} A_{\rm CURN} = -14.28^{+0.21}_{-0.21}$, with a Bayes factor $\ln\mathcal{B}=4.46$ for \ac{CURN}. Given its large pulsar number, high cadence, and southern hemisphere location, the \ac{MPTA} data set is promising for testing \ac{GR}, including constraints on the graviton mass via the modified dispersion relation and the generalised correlation function.

Looking further ahead, the upcoming \ac{SKA}--\ac{PTA} \cite{Janssen:2014dka,Weltman:2018zrl,SKAOPulsarScienceWorkingGroup:2025oyu} will deliver a transformative leap in sensitivity, and the \ac{MPTA} serves as its critical pathfinder. MeerKAT is the South African \ac{SKA} precursor telescope, located at the future site of \ac{SKA}-Mid, and will eventually be embedded into the full \ac{SKA}-Mid array \cite{Bailes:2020qai,Spiewak:2022btk}. As the most sensitive centimetre-wavelength radio telescope in the southern hemisphere, MeerKAT serves both as a pathfinder for \ac{SKA}-Mid and as an immediate science instrument. The \ac{SKA} is expected to be a factor of three to four more sensitive than any other southern hemisphere facility, with the \ac{SKA}--\ac{PTA} approaching strain amplitudes of $6\times10^{-16}$ or better at $f\sim1\,{\rm yr}^{-1}$, at least one order of magnitude improvement over current \acp{PTA} \cite{Lazio:2013mea,SKAOPulsarScienceWorkingGroup:2025oyu}. With the ability to time up to 174 spin-stable millisecond pulsars~\cite{SKAOPulsarScienceWorkingGroup:2025oyu}, the \ac{SKA}--\ac{PTA} has a dramatically enhancing sensitivity to tests of fundamental physics, including constraints on the graviton mass \cite{Weltman:2018zrl}. The \ac{MPTA} thus represents an important stepping stone towards the \ac{SKA} era, providing both immediate science and critical experience for optimising future \ac{SKA}--\ac{PTA} observing strategies.

In this work, we use the latest \ac{MPTA} 4.5-year data release \cite{Miles:2024seg} to derive upper limits on the graviton mass $\mg$, and we perform forecasts for the \ac{SKA}--\ac{PTA} \cite{SKAOPulsarScienceWorkingGroup:2025oyu} to assess its anticipated sensitivity. We model the effects of massive graviton on both the \ac{GW} dispersion relation and the \ac{PTA} correlation function, and perform Bayesian inference to constrain $\mg$. The \ac{MPTA}'s high cadence and large pulsar count offer a valuable opportunity to tighten existing bounds and to test possible departures from the massless-graviton limit explored in literature. In addition, our forecasts for the upcoming \ac{SKA}--\ac{PTA} provide a forward-looking perspective on how future facilities will improve upon the bounds derived from current \acp{PTA}.

The remainder of this paper is organised as follows. Section~\ref{sec:theory} presents the theoretical framework, including the modified dispersion relation and the generalised correlation curve. Section~\ref{sec:data-method} describes the \ac{MPTA} data products and the Bayesian inference pipeline. Section~\ref{sec:results} reports our main results and upper limits across the three datasets. Section~\ref{sec:ska-forecast} presents forecasts for \ac{SKA}--\ac{PTA} to assess its anticipated sensitivity to the graviton mass. Finally, Section~\ref{sec:discussion-conclusion} discusses our findings, examines systematic effects, and concludes with future prospects.

\section{Correlation curve for a massive graviton}\label{sec:theory}

The presence of a non-zero graviton mass modifies the propagation of \acp{GW} at a fundamental level. While a full field-theoretic description can be formulated via a Fierz–Pauli mass term in the Lagrangian \cite{Fierz:1939ix,deRham:2014zqa}, a simpler and more direct phenomenological approach is to adopt the modified dispersion relation for massive \acp{GW}, i.e.,
\begin{equation}
E^2 = p^2 c^2 + \mg^2 c^4\,, \label{eq:dispersion}
\end{equation}
where $E$ and $p$, respectively, are the energy and momentum of the graviton, $\mg$ is the graviton mass, and $c$ is the speed of light. For \acp{GW} with angular frequency $\omega = 2\pi f$, the energy is $E = \hbar \omega$ and the momentum is $p = \hbar k$, with $k$ being the wavenumber. The dispersion relation gives the group velocity $v_{g}=d\omega/dk$ as a function of \ac{GW} frequency 
\begin{equation}
v_{g} = c \, \sqrt{1 - \frac{\mg^2 c^4}{h^2 f^2}} \approx c \left(1 - \frac{\mg^2 c^4}{2 h^2 f^2}\right) \quad \text{for} \quad f \gg \frac{\mg c^2}{h}\,, \label{eq:group_velocity}
\end{equation}
where $h = 2\pi\hbar$ is Planck's constant. To probe graviton mass, the group velocity relative to the speed of light is the key observational signature for \acp{PTA}.

A \ac{SGWB} induces correlated pulse-arrival-time residuals between a pair of pulsars. For massive tensor modes, the angular correlation is expanded as \cite{Bernardo:2022rif}
\begin{equation}
    \gamma(\zeta_{ab}) = \sum_{\ell} \frac{2\ell+1}{4\pi} C_{\ell} P_{\ell}(\cos\zeta_{ab})\,,
\end{equation}
where indices $a$ and $b$ label the two pulsars, $\zeta_{ab}$ is their angular separation, and $P_{\ell}$ are Legendre polynomials. The coefficient $C_{\ell}$ takes the form
\begin{equation}
    C_{\ell}
    =
    \frac{1}{\sqrt{\pi}}
    J_{\ell}\left(\frac{v_g}{c},\frac{fD_a}{c}\right)
    J_{\ell}^{\ast}\left(\frac{v_g}{c},\frac{fD_b}{c}\right)\,,
\end{equation}
where $D_a$ and $D_b$ are the distances to pulsars $a$ and $b$, respectively, and
\begin{equation}
    J_{\ell}\left(\frac{v_g}{c},\frac{fD}{c}\right)
    =
    \sqrt{2}\pi i^{\ell}
    \sqrt{\frac{(\ell+2)!}{(\ell-2)!}}
    \int_{0}^{2\pi (fD/c)(v_g/c)}
    \frac{dx}{v_g/c}
    e^{i x/(v_g/c)}
    \frac{j_{\ell}(x)}{x^{2}}\,.
\end{equation}
with $j_{\ell}(x)$ representing the spherical Bessel function of order $\ell$. For each fixed $\ell$, the large-distance limit $fD/c\gg1$ sends the upper integration bound to infinity. In the additional luminal limit $v_g/c\to1$, the integral gives \cite{Bernardo:2022rif}
\begin{equation}
\int_0^\infty dx\,e^{ix}\frac{j_\ell(x)}{x^2}
=
2i^{\ell-1}\frac{(\ell-2)!}{(\ell+2)!},
\qquad
C_\ell^{\rm HD}
=
8\pi^{3/2}\frac{(\ell-2)!}{(\ell+2)!},
\quad \ell\geq2.
\label{eq:hd-multipoles}
\end{equation}
Thus the pulsar distances drop out of the multipoles in this limit.
Following Ref.~\cite{Bernardo:2022rif}, we convert $\gamma(\zeta_{ab})$ into the correlation function for massive gravitons, denoted $\Gamma(\zeta_{ab};v_{g})$, by imposing the normalisation $\Gamma(\zeta_{ab}=0^+;v_{g}=c)=0.5$.
 {With this normalisation, the multipole sum gives the standard HD cross-correlation for distinct pulsars:
\begin{equation}
\Gamma_{\rm HD}(\zeta)
=\frac{3}{2}\sum_{\ell=2}^{\infty}
\frac{(2\ell+1)(\ell-2)!}{(\ell+2)!}P_\ell(\cos\zeta)
=\frac12-\frac{u}{4}+\frac32u\ln u
\qquad \left(u=\frac{1-\cos\zeta}{2}\right).
\label{eq:hd-distance-limit}
\end{equation}
The standard HD correlation is recovered in the joint large-distance $\frac{fD}{c}\gg1$ and luminal limit and $v_g\to c$.}
In this work, we use \texttt{PTAfast} \cite{Bernardo:2022rif,Bernardo:2023mxc} to produce the above theoretical predictions for a massive graviton.

\begin{figure*}[t]
\centering
 \includegraphics[width=\textwidth]{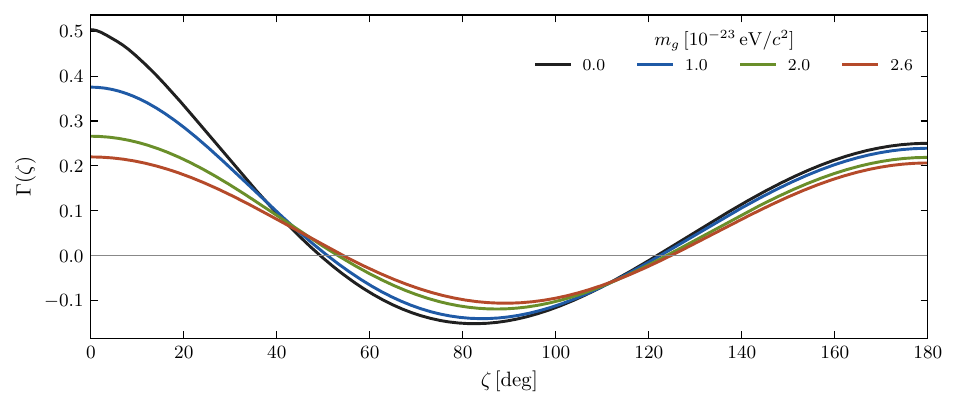}
\caption{Impact of a non-zero graviton mass on the correlation curve. The black curve represents the standard \ac{HD} correlation expected in \ac{GR}. The coloured curves show the modified tensor correlations for $\mg=1.0$, $2.0$, and $2.6$ in units of $10^{-23}\,\mathrm{eV}/c^{2}$.}
\label{fig:HD_correction}
\end{figure*}

Fig.~\ref{fig:HD_correction} illustrates impacts of a non-zero graviton mass on the \ac{HD} correlation curve. The standard \ac{HD} curve is shown as a black line. For a fixed \ac{GW} frequency within the \ac{PTA} band, increasing the graviton mass introduces a systematic deformation of the correlation pattern. The amplitude is enhanced for intermediate angular separations, while it is suppressed at both small and large angular scales. The curves shown in Fig.~\ref{fig:HD_correction} span the mass range relevant to the present \ac{MPTA} posterior support. Consequently, a precise measurement of the modified correlation offers a powerful means for \acp{PTA} to constrain the graviton mass, and in principle, to uncover evidence for a non-zero $\mg$.

\section{Bayesian analysis of \ac{MPTA} data}
\label{sec:data-method}

Our analysis is based on the angular-correlation products released by the \ac{MPTA} Collaboration \cite{Miles:2024seg}, which provide cross-correlation measurements across 15 angular bins constructed from 3403 pulsar pairs during an observing duration $t_{\rm obs}=4.5$\,years. To account for the potential impact of intrinsic pulsar noise on the inferred \ac{SGWB} signature, the \ac{MPTA} analysis adopts three distinct noise configurations, designated DATA, ER, and ALT. The DATA model follows a data-driven approach, retaining only those noise processes that are statistically supported by the single-pulsar timing solutions, and comprises 390 free parameters. The ER model adopts a more conservative strategy by including an achromatic red-noise component for every pulsar irrespective of its individual Bayesian evidence, thereby expanding the parameter space to 532 degrees of freedom. The ALT model is a slight variant of ER, differing only in the removal of an anomalously steep achromatic red-noise term associated with the pulsar PSR\,J2129-5721, which reduces the parameter count to 530.

These three configurations serve as a systematic test of the sensitivity of \ac{HD} correlation recovery to noise model assumptions \cite{Miles:2024seg}. Under the fixed-parameter solutions derived from the Bayesian posterior maxima, the optimal statistic yields systematically varying cross-correlation amplitudes. Specifically, the DATA model gives 
\begin{equation}
{A}_{\mathrm{DATA}}^2 = \left(5.65\pm1.20\right)\times10^{-29}\,,\label{eq:a2data}
\end{equation}
the ALT model gives the highest value at 
\begin{equation}
{A}_{\mathrm{ALT}}^2 = (7.10\pm1.80)\times10^{-29}\,,\label{eq:a2alt}
\end{equation}
while the ER model returns a substantially lower amplitude of 
\begin{equation}
{A}_{\mathrm{ER}}^2 = (3.70\pm1.90)\times10^{-29}\,.\label{eq:a2er}
\end{equation}
This gradient clearly demonstrates that the anomalous red-noise component in PSR\,J2129-5721, which is present in ER but absent in ALT, can effectively absorb power that would otherwise contribute to the global correlated signal. In terms of frequentist significance, both the DATA and ALT models yield false-alarm probabilities corresponding to approximately $3.2$--$3.4\sigma$ when evaluated via sky and phase scrambling, whereas the ER model fails to produce any significant evidence for \ac{HD} correlations.

For the purpose of this study, we adopt the published optimal-statistic tensor-power measurements derived from the three noise models as truncated Gaussian priors on the non-negative tensor power, following the \ac{MPTA} Collaboration's prescription. Specifically, we use the aforementioned amplitude measurements $A^2_{\rm DATA}$, $A^2_{\rm ER}$, and $A^2_{\rm ALT}$ from the \ac{MPTA} analysis as the central values and their corresponding uncertainties as the $1\sigma$ errors.
Therefore, a likelihood function is constructed from the angular correlation measurements, assuming Gaussian noise statistics. It is explicitly given as 
\begin{equation}
 \ln{\cal L}=-\frac12\sum_{i=1}^{15} \frac{(\mu_{i}^{\mathrm{obs}}-\mu_{i}^{\mathrm{th}})^2}{\sigma_{i}^{2}}\,,
 \label{eq:likelihood}
\end{equation}
where the index $i$ labels the angular-separation bin, and we introduce a theoretical prediction 
\begin{equation}
    \mu_{i}^{\mathrm{th}} = A^{2}\Gamma(\zeta_{i};v_{g})\,,\label{eq:nuisance-mean-prime}
\end{equation}
with $\mu_{i}^{\mathrm{obs}}$ representing the observational central value and $\sigma_{i}$ the corresponding $1\sigma$ \ac{CL} observational error. 
Here, due to the absence of the full covariance matrix in the released products, we adopt a diagonal Gaussian likelihood. This simplification is conservative in the sense that it neglects off-diagonal correlations among angular bins, which, if properly accounted for, would likely lead to tighter constraints on the model parameters. 
{We also consider \ac{CURN} as an additional source of error by adding $\Sigma^2$ to the variance of each angular bin, i.e., $\sigma_i^2\rightarrow\sigma_i^2+\Sigma^2$, where $\Sigma$ characterises the \ac{CURN}. The corresponding likelihood is
\begin{equation}
 \ln {\cal L}_{\rm CURN}=-\frac{1}{2}\sum_{i=1}^{15}
 \left[
 \frac{(\mu_i^{\rm obs}-\mu_i^{\rm th})^2}{\sigma_i^2+\Sigma^2}
 +\ln\!\left(2\pi(\sigma_i^2+\Sigma^2)\right)
 \right].
 \label{eq:likelihood-curn}
\end{equation}
We adopt a uniform prior $\Sigma\sim{\cal U}(0,10^{-28})$.
}

The Bayesian parameter inference is implemented using \texttt{Cobaya} \cite{Torrado:2020dgo}, a \ac{MCMC} sampler. The angular-correlation measurements are compressed over frequency. Accordingly, we adopt the lowest resolvable Fourier frequency, $f_{\rm ref}=1/t_{\rm obs}$, to map $m_g$ onto $v_g$. This choice reflects the low-frequency dominance of the pulsar-timing residual spectrum and the scaling $1-v_g/c\propto m_g^2/f^2$. The requirement of a real propagation speed then yields the kinematic bound $m_g\leq h f_{\rm ref}/c^2$, which sets the upper limit of the uniform prior on $m_g$. We therefore adopt $0\leq m_g\leq h f_{\rm ref}/c^2$; a uniform prior on $m_g$ is equivalent to a LinearExp prior on $\log_{10}[m_g/({\rm eV}/c^2)]$. Convergence is monitored via the Gelman--Rubin diagnostic, ensuring that all chains achieve $R-1 < 0.001$. We conservatively discard the first 50\% of each chain for burn-in.

Beyond the \ac{SGWB}, which carries a characteristic quadrupolar signature, we consider two further common signals exhibiting distinct orientational dependencies. Clock inaccuracies introduce a monopole, whereas ephemeris defects produce a dipole \cite{Tiburzi:2015kqa,Vigeland:2018ipb}. To impose rigorous upper bounds on $m_{g}$, we compare the measured correlation curve against a linear superposition of all three contributions. This combined template is expressed as
\begin{equation}
 \mu_{i}^{\mathrm{th}} = A^{2} \Gamma(\zeta_i;v_{g}) + M + D \cos\zeta_i \,, \label{eq:nuisance-mean}
\end{equation}
where $M$ and $D$ stand for monopole and dipole coefficients, respectively. We take uniform priors of
${\cal U}(-10^{-28},10^{-28})$ for both $M$ and $D$. To assess the impact of additional angular components, we compare two inference configurations, i.e., one using Eq.~(\ref{eq:nuisance-mean-prime}) and the other incorporating all three correlations simultaneously, as shown in Eq.~(\ref{eq:nuisance-mean}). From the same observational data set, we extract the tensor parameters under each scheme and quantify the discrepancies between the two outcomes. This contrast reveals whether the monopolar and dipolar contributions materially alter the recovered tensor signal.

\section{Upper bounds on \texorpdfstring{$\mg$}{mg} from \ac{MPTA}}
\label{sec:results}

\begin{figure*}[t]
\centering
\includegraphics[width=\textwidth]{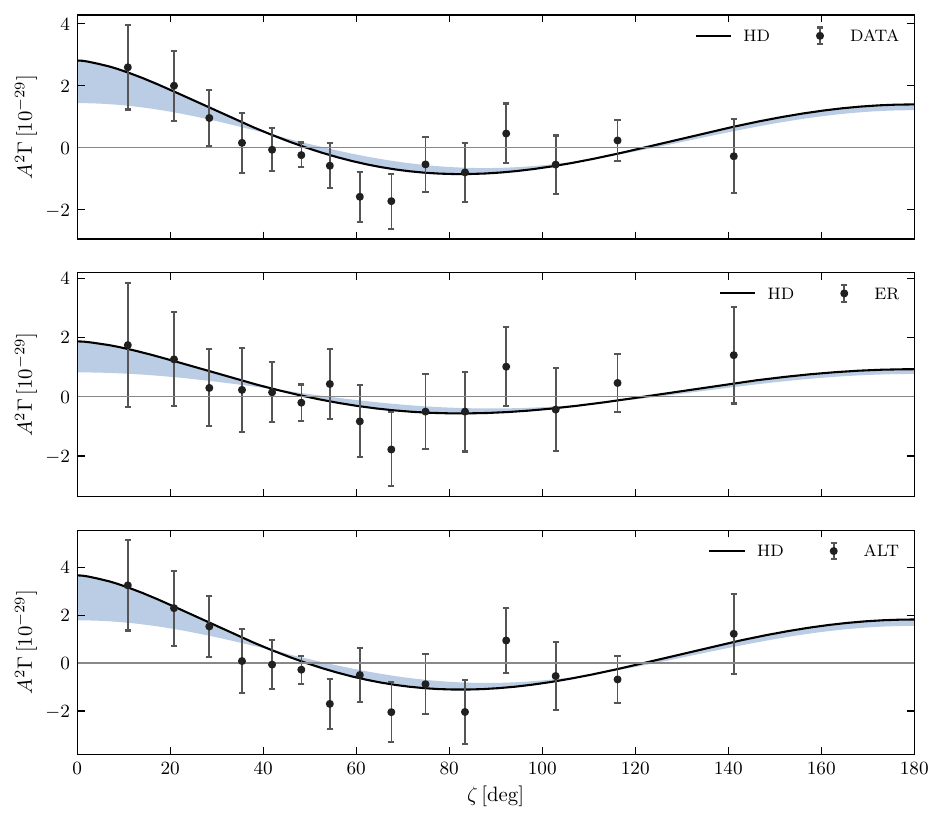}
\caption{
Angular-correlation measurements and massive-graviton allowed regions inferred from the \ac{MPTA} observations. The points with error bars are the published angular-correlation measurements. The shaded bands show the range of the tensor template obtained by varying $\mg$ from zero to its 90\% posterior upper limit while fixing $A^2$ to its posterior median. The solid black curves show the massless \ac{HD} template with the same $A^2$ normalisation. The panels, ordered from top to bottom, correspond to DATA, ER, and ALT.}
\label{fig:angular}
\end{figure*}

In Figure~\ref{fig:angular}, we display the angular-correlation data together with the shaded regions allowed by the 90\% upper limit on $\mg$. The latter show how far the tensor \ac{HD} curve can be deformed by a massive graviton within the posterior support of each data set.
The solid \ac{HD} curves, overlaid for reference, represent the fixed theoretical expectation for a massless graviton. 

Comparison between the standard \ac{HD} curve and the inferred intervals reveals that the 90\% credible regions comfortably encompass the \ac{HD} reference for all three \ac{MPTA} data sets, indicating full consistency with \ac{GR}. However, the widths of the allowed intervals differ appreciably among the three data sets, with some yielding tighter constraints than others. These variations trace the different intrinsic-noise prescriptions used in the DATA, ER, and ALT reductions. The observed scatter in interval widths therefore highlights the importance of the noise treatment, but does not undermine the overall finding that the current observations remain compatible with the massless-graviton hypothesis.

\begin{figure*}[t]
\centering
\includegraphics[width=\textwidth]{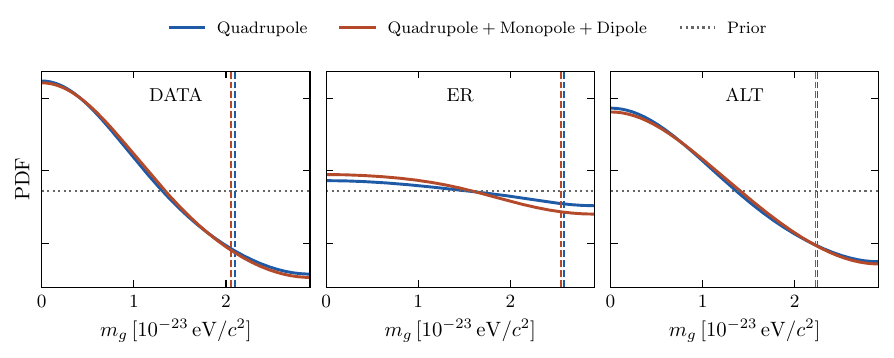}
\caption{\label{fig:posteriors}
One-dimensional marginalised posterior distributions for $\mg$, derived from the three \ac{MPTA} data sets, i.e., DATA, ER, and ALT, for the baseline model and the extended model including monopolar and dipolar correlations. For each distribution, the vertical dashed lines mark the corresponding 90\% credible upper limits, providing a quantitative complement to the shaded-band inference presented in Figure~\ref{fig:angular}. The posterior peaks for all three data sets remain compatible with the massless-graviton expectation, while the widths of the distributions vary across data sets, reflecting differences in the constraining power of each reduction. The horizontal dotted lines denote the priors adopted in this work.}
\end{figure*}

To complement the inference results presented in Figure~\ref{fig:angular}, we turn to Figure~\ref{fig:posteriors}, which displays the one-dimensional marginalised posterior distributions for $\mg$ obtained from the three \ac{MPTA} data sets. For each data set and model, the vertical dashed line marks the 90\% upper limit, providing a complementary visual summary alongside the shaded-band representation already discussed. For comparison, the horizontal dotted lines denote the priors adopted in this work. 

Consistent with the interval comparisons in Figure~\ref{fig:angular}, the posterior peaks for all three data sets remain fully compatible with the massless-graviton limit. The widths of the posterior distributions, however, vary dramatically across the data sets. Among the three reductions, DATA yields the tightest, most sharply peaked posterior, indicating the strongest constraining power on $\mg$. ALT provides a moderately broader distribution, while ER exhibits an extremely broad, nearly flat posterior, implying that it offers little to no meaningful constraint on $m_{g}$.

This hierarchy in constraining power can be understood from the distinct intrinsic-noise modelling choices adopted by each reduction \cite{Miles:2024seg}. The DATA model employs a parsimonious, data-driven noise selection, whereas the ER model conservatively includes achromatic red noise for every pulsar, including an anomalous component in PSR\,J2129-5721, substantially expanding the parameter space. The ALT model removes this specific anomalous term while retaining red noise for other pulsars. As demonstrated in Ref.~\cite{Miles:2024seg}, this anomalous component exhibits a pronounced degeneracy with a global \ac{GW} signal, such that its presence in ER absorbs significant power that would otherwise contribute to the correlated \ac{HD} signature, thereby severely diluting the sensitivity to $\mg$. In contrast, the more parsimonious noise budget in DATA allows the quadrupolar correlation to be extracted with greater statistical significance. The intermediate behaviour of ALT, which excludes only the most problematic noise term, places it between these two extremes. Despite these pronounced differences in constraining power, the central values of the posteriors show no systematic shifts across data sets, reinforcing the earlier finding that the quadrupole extraction is not biased by the inclusion of additional angular components.

For quantitative reference, the 90\% credible upper limits on $\mg$ derived from each data set are as follows:
\begin{align}
\mg&<2.10\times10^{-23}\,{\rm eV}/c^2\quad({\rm DATA})\,, \label{eq:mg_DATA_Q} \\
\mg&<2.58\times10^{-23}\,{\rm eV}/c^2\quad({\rm ER})\,,   \label{eq:mg_ER_Q} \\
\mg&<2.25\times10^{-23}\,{\rm eV}/c^2\quad({\rm ALT})\,.  \label{eq:mg_ALT_Q}
\end{align}
When the extended model, which incorporates monopolar and dipolar contributions, is adopted, the corresponding limits become
\begin{align}
\mg&<2.06\times10^{-23}\,{\rm eV}/c^2\quad({\rm DATA})\,, \label{eq:mg_DATA_E} \\
\mg&<2.55\times10^{-23}\,{\rm eV}/c^2\quad({\rm ER})\,,   \label{eq:mg_ER_E} \\
\mg&<2.22\times10^{-23}\,{\rm eV}/c^2\quad({\rm ALT})\,.  \label{eq:mg_ALT_E}
\end{align}
Comparing the two sets of values, we find that the inclusion of additional angular correlations leads to only minor adjustments in the numerical bounds, with the shifts remaining well within the uncertainties of the inference. This quantitative consistency, together with the visual evidence from both figures, reinforces our earlier conclusion drawn from Figure~\ref{fig:angular} that the inclusion of monopolar and dipolar contributions does not materially alter the mass inference. The current \ac{MPTA} observations, across all three reductions and both modelling schemes, remain fully compatible with the massless-graviton expectation, indicating full consistency with \ac{GR}.

This comparison goes beyond a generic statement that ``noise matters''. The DATA, ER, and ALT configurations share identical telescope, observing cadence, pulsar set, timing baseline, and angular-correlation estimator. The only documented difference between ER and ALT is the treatment of the achromatic red-noise process associated with PSR\,J2129-5721 \cite{Miles:2024seg}. Removing this single component in ALT brings the inferred mass posterior into close agreement with that obtained from DATA. This behaviour identifies a concrete physical pathway through which intrinsic-noise modelling can influence a gravity constraint. Low-frequency power that could contribute to the common cross-correlated sector is instead absorbed by the pulsar-specific red-noise terms, effectively diluting the angular-correlation signature. The effect is further amplified by the favourable sky geometry of the highest-precision pulsars, which enhances the leverage of the quadrupolar template.

The consistency between ALT and DATA is physically encouraging. The preference for small graviton masses does not hinge on a single reduction scheme. It re-emerges when the conservative ER model is modified in the direction suggested by the \ac{MPTA} residuals. While this does not establish ALT as the unique correct description of the noise in PSR\,J2129-5721, it demonstrates that the preference for the massless limit is tied to the angular-correlation structure that the \ac{MPTA} Collaboration identified as less susceptible to this particular modelling ambiguity. Taken together, these findings suggest that the quadrupolar signal, and hence the resulting upper limits on $\mg$, survives a meaningful subset of noise-model variations, reinforcing the reliability of the constraints presented here.

Having established that the inclusion of monopolar and dipolar components does not qualitatively alter the mass inference, we now examine the fitted values of the additional independent parameters $M$ and $D$ under the extended model. The 68\% credible intervals for both parameters, obtained from each of the three \ac{MPTA} data sets, are as follows
\begin{align}
M_{\mathrm{DATA}} &= (-0.681 \pm 2.77) \times 10^{-30}\,,  \\
D_{\mathrm{DATA}} &= (-4.10  \pm 4.69) \times 10^{-30}\,;  \\
M_{\mathrm{ER}}   &= (\phantom{-}2.53  \pm 4.03) \times 10^{-30}\,,  \\
D_{\mathrm{ER}}   &= (-7.09  \pm 6.79) \times 10^{-30}\,;  \\
M_{\mathrm{ALT}}  &= (-1.63  \pm 4.00) \times 10^{-30}\,,  \\
D_{\mathrm{ALT}}  &= (-4.19  \pm 6.78) \times 10^{-30}\,.
\end{align}
These intervals show no coherent evidence for a monopolar or dipolar contribution across the three reductions, supporting the interpretation that the graviton-mass constraint is driven by the quadrupolar tensor correlation rather than by these additional angular templates.

Most of the 68\% intervals overlap zero. The exception is the dipole coefficient in the ER reduction, which shows a mild negative offset at this credibility level. Because this offset is not reproduced by DATA or ALT, it does not constitute coherent evidence for a dipolar contribution across the \ac{MPTA} reductions. The uncertainties on $M$ and $D$ are considerably larger for the ER data set than for DATA and ALT, reflecting the degraded constraining power of the ER model due to the inflated noise budget discussed earlier. In the DATA and ALT reductions, the inferred monopole and dipole amplitudes remain close to zero, with typical scales of order $10^{-30}$, well below the tensor power $A^2 \sim 10^{-29}$. This suggests that the clock and ephemeris systematics that would give rise to these angular patterns are either absent or effectively marginalised out in the current analysis.   

The absence of a coherent nonzero $M$ or $D$ across reductions carries two important implications. First, it demonstrates that the extended model does not systematically absorb tensor power into the monopole or dipole channels. Second, it reinforces the robustness of the graviton-mass constraints presented in the previous sections. That is, the tensor correlation is not strongly degenerate with the lower-order angular correlations, and the upper limits on $\mg$ are stable against the inclusion of these extra components. We therefore conclude that the extended model serves as a successful consistency check, rather than a necessary improvement, for the current \ac{MPTA} data.

Table~\ref{tab:curn-evidence} summarizes the results obtained with the additional \ac{CURN} contribution introduced in Eq.~\eqref{eq:likelihood-curn}. The DATA, ER, and ALT mass upper limits become $(\DATACurnMass,\ERCurnMass,\ALTCurnMass)\times10^{-23}\,{\rm eV}/c^2$, respectively, corresponding to increases of $(\DATACurnChange,\ERCurnChange,\ALTCurnChange)\%$ relative to the baseline results. Thus, the inclusion of \ac{CURN} has only a minor effect on the inferred graviton-mass limits.

To distinguish an upper limit from evidence for a nonzero mass, we compare a massive-graviton model, ${\cal M}_{m}$, with the massless model, ${\cal M}_{0}$, and define
\begin{equation}
\ln B_{m0}=\ln Z({\cal M}_{m})-\ln Z({\cal M}_{0}).
\label{eq:mass-bayes-factor}
\end{equation}
Thus, positive values favour a freely varying graviton mass and negative values favour the massless model. The baseline and \ac{CURN}-extended log Bayes factors are reported in Table~\ref{tab:curn-evidence}; none of the three reductions provides significant evidence for a nonzero graviton mass.

\begin{table}[t]
\centering
{
\caption{
Constraints on the graviton mass for the three MPTA noise models, 
with and without \ac{CURN}. 
The quoted values of $\Sigma$ are the posterior median and central 68\% credible interval, 
while $\Sigma^{90\%}$ and $m_g^{90\%}$ denote one-sided 90\% credible upper limits. 
$\ln B_{m0}$ denotes the log Bayes factor between the massive- and massless-graviton models. 
Masses are in units of $10^{-23}\,{\rm eV}/c^2$ and $\Sigma$ in units of $10^{-29}$.
}
\label{tab:curn-evidence}
\small
\begin{adjustbox}{max width=\linewidth}
\begin{tabular}{lcccccc}
\toprule
& \multicolumn{2}{c}{CURN}
& \multicolumn{2}{c}{$m_g^{90\%}$}
& \multicolumn{2}{c}{$\ln B_{m0}$} \\
\cmidrule(lr){2-3}
\cmidrule(lr){4-5}
\cmidrule(lr){6-7}
Noise model
& $\Sigma$
& $\Sigma^{90\%}$
& Baseline
& CURN
& Baseline
& CURN \\
\midrule
DATA & $0.23^{+0.26}_{-0.16}$ & 0.580 & 2.10 & 2.20 & -0.73 & -0.64 \\
ER   & $0.26^{+0.30}_{-0.18}$ & 0.674 & 2.58 & 2.59 & -0.09 & -0.08 \\
ALT  & $0.31^{+0.35}_{-0.22}$ & 0.782 & 2.25 & 2.31 & -0.58 & -0.51 \\
\bottomrule
\end{tabular}
\end{adjustbox}
}
\end{table}
\section{Forecast for \ac{SKA}--\ac{PTA} sensitivity to $m_{g}$}
\label{sec:ska-forecast}

The preceding \ac{MPTA} analysis has established current constraints on $\mg$. As MeerKAT is a pathfinder for the \ac{SKA}, we assess how this inference would translate to future \ac{SKA}--\ac{PTA} data products, assuming the measured correlation remains centred on the \ac{HD} curve but with substantially smaller uncertainties. The \ac{SKA} is expected to provide a major gain for pulsar timing and fundamental-physics tests through a larger high-precision pulsar sample, improved timing sensitivity, and long observing baselines \cite{Janssen:2014dka,Weltman:2018zrl,SKAOPulsarScienceWorkingGroup:2025oyu}. We therefore expect it to place stronger constraints on $m_{g}$, further testing the \ac{GR}. 

To quantify the \ac{SKA}--\ac{PTA} sensitivity to the graviton mass, we present three forecasts with observation baselines of $T_{\rm obs}=10$, $20$, and $50$\,years.
For each observing span, we use a simulated catalogue containing 174 pulsars with sky positions drawn isotropically~\cite{SKAOPulsarScienceWorkingGroup:2025oyu}. All pulsars are assigned the same observing cadence, $\Delta t=14\,{\rm days}$, and the same white-noise rms timing uncertainty, $\sigma_a=0.1\,\mu{\rm s}$ \cite{Babak:2024yhu}. Both of the red-noise amplitude $\log_{10} A_a^{\rm RN}$ and the dispersion-measure amplitude $\log_{10} A_a^{\rm DM}$ are independently drawn from ${\cal U}(-17,-13)$, while the corresponding spectral indices $\gamma_{a}^{\rm RN}$ and $\gamma_{a}^{\rm DM}$ are drawn from ${\cal U}(1,5)$ \cite{Babak:2024yhu}. 
We emphasize that these forecasts are idealized, as the actual \ac{SKA}--\ac{PTA} sensitivity will depend on pulsar jitter and on the distribution of intrinsic red-noise properties across the pulsar population \cite{Janssen:2014dka,SKAOPulsarScienceWorkingGroup:2025oyu}.

The forecast correlation errors are obtained from a Fisher reconstruction of the \ac{HD} correlation curve, following the methodology of Ref.~\cite{Babak:2024yhu}. 
If the fitted independent parameters are denoted by $\theta_\alpha$, the Fisher matrix is written as
\begin{equation}
F_{\alpha \beta} = \sum_{f_k} \sum_{a,b,c,d} C_{ab}^{-1} C_{cd}^{-1}
\frac{\partial\left(R_{bc} S_h\right)}{\partial \theta^\alpha}
\frac{\partial\left(R_{da} S_h\right)}{\partial \theta^\beta} \,,\label{eq:fisher}
\end{equation}
where $f_k=k/T_{\rm obs}$ labels the $k$-th Fourier-frequency bin, with $k=1,2,...,30$, $R_{ab}$ is the angular
response matrix, $S_h$ is the \ac{GW} strain power spectrum, and
$C_{ab}$ includes both the pulsar-noise and signal contributions, i.e.,
\begin{equation}
C_{ab}=\delta_{ab} P_{n,a}+R_{ab} S_h(f) \,.
\end{equation}
To assess the impact of \ac{CV}, we compare the fiducial forecast using the full covariance with a noise-only approximation, $C_{ab}=\delta_{ab}P_{n,a}$. The full covariance includes both stochastic signal variance, including \ac{CV}, and signal--noise terms, whereas both contributions are absent in the noise-only case. We refer to these two cases as with and without \ac{CV}, respectively.

Here, indices such as $a$ and $b$ enumerate the pulsars, ranging up to the total number of pulsars. 
In the fiducial model, we use the \ac{HD} correlation $\Gamma(\zeta_{ab})$ in the formula of $R_{ab}$. 
The pulsar-noise power $P_{n,a}$ is a sum of white noise, red noise, and a
dispersion-measure component. The white-noise term is
\begin{equation}
P_a^{\rm WN}=2\sigma_{a}^2\Delta t\,,
\end{equation}
where $\Delta t$ and $\sigma_a$ are fixed to the values specified above. The red noise is given by a power law,
\begin{equation}
P_a^{\rm RN}(f)=\frac{(A_a^{\rm RN})^2}{12\pi^2 f_{\rm yr}^3}\left(\frac{f_{\rm yr}}{f}\right)^{\gamma_a^{\rm RN}} \,,
\end{equation}
where $f_{\rm yr}=1/(1\,{\rm year})$ stands for a reference frequency. The dispersion-measure contribution is treated as an additional power law,
\begin{equation}
P_a^{\rm DM}(f)
=
\frac{(A_a^{\rm DM})^2}{12\pi^2 f_{\rm yr}^3}
\left(\frac{f_{\rm yr}}{f}\right)^{\gamma_a^{\rm DM}}\,,
\end{equation}
with the random assignments stated above.

Once the Fisher matrix is given, the Cramer--Rao bound provides the $1\sigma$ measurement uncertainty of $\theta_{\alpha}$, i.e., 
\begin{equation}
    \sigma_{\theta_\alpha} = \sqrt{ \left(F^{-1}\right)_{\alpha\alpha} }\,,
\end{equation}
where $F^{-1}$ is the inverse of the Fisher matrix. 
Here, we infer the following set of 17 model parameters
\begin{equation}
\theta_\alpha =
\left\{\alpha_{\rm PL}, n_{\rm T}, b_1,\ldots,b_{15}\right\}\,.
\end{equation}
The first two parameters stand for the amplitude and tilt of the fiducial
power-law \ac{SGWB} used to define the covariance and its
derivatives, namely, $\log_{10}S_{h}(f)\sim\log_{10}\alpha_{\rm PL}+(n_{\rm T}-3)\log_{10}(f/f_{\rm yr})$. 
The remaining parameters $b_i$ describe a piecewise-constant angular correlation function in 15 equal bins over $0\leq\zeta\leq\pi$. Thus $b_i$ is constant in the interval from $(i-1)\pi/15$ to $i\pi/15$, with $i=1,\ldots,15$. In this study, we use the \texttt{fastPTA} \cite{Babak:2024yhu} package to perform the above Fisher forecast.

\begin{figure*}[!t]
\centering
\includegraphics[width=\textwidth]{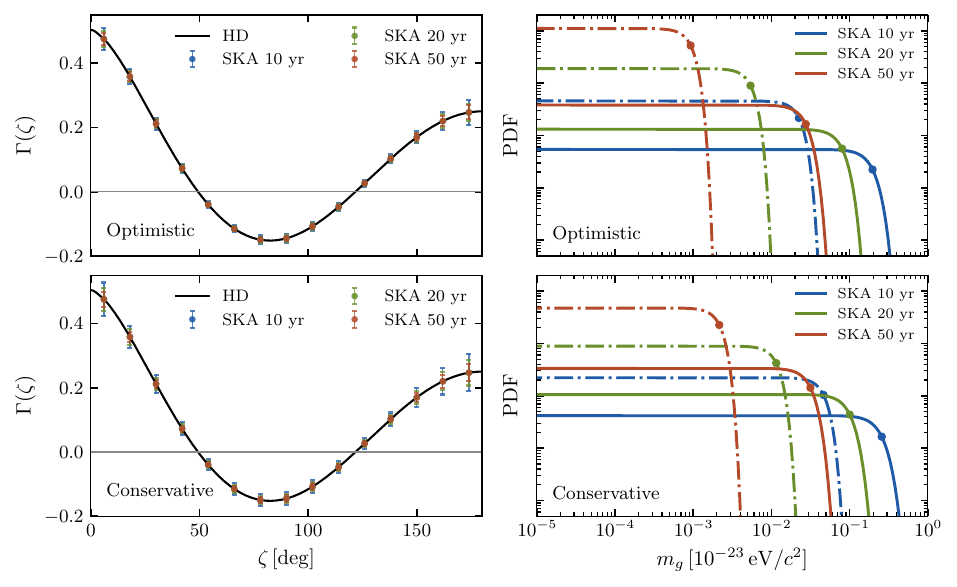}
\caption{\label{fig:ska-forecast}
Projected \ac{SKA}--\ac{PTA} angular-correlation precision (left panels) and sensitivity to measuring $m_{g}$ (right panels). The upper two panels assume 174 pulsars with an rms timing uncertainty of $0.1\,{\rm \mu s}$ for every pulsar (optimistic), whereas the lower two panels assume 50 pulsars timed at $0.1\,{\rm \mu s}$ and the remaining 124 at $0.5\,{\rm \mu s}$ (conservative). In the left panels, the coloured points and $1\sigma$ error bars show the 10-, 20-, and 50-year forecasts centred on the fiducial \ac{HD} curve, with the error bars including stochastic signal variance. The right panels show the corresponding normalized one-dimensional posteriors for $\mg$, with \ac{CV} (solid) and without \ac{CV} (dash-dotted), as defined in the text. The dots mark their one-sided 90\% credible upper limits.}
\end{figure*}

The forecast errors on the correlation, denoted as $\sigma_{b_{i}}$ hereafter, are derived from the uncertainties in the \(b_i\). 
They are depicted explicitly in the left panel of Fig.~\ref{fig:ska-forecast}, where the deep blue, olive green, and burnt orange error bars correspond to the \ac{SKA}--\ac{PTA} sensitivities for observational baselines of 10, 20, and 50\,years, respectively. As expected, $\sigma_{b_i}$ decreases progressively as the observational baseline increases. For reference, the solid black curve in the same panel represents the standard \ac{HD} correlation, which serves as our fiducial model.

To derive prospective constraints on \(m_g\) from the forecasted \ac{SKA}--\ac{PTA} sensitivities, we construct a likelihood function of the form
\begin{equation}
\ln {\cal L} = -\frac{1}{2}\sum_{i=1}^{15} \frac{\left(\Gamma(\zeta_i;c)-\Gamma(\zeta_i;v_{g})\right)^2}{\sigma_{b_i}^2}\,,
\end{equation}
where the sum runs over angular bins, \(\Gamma(\zeta_i;c)\) are the {mean correlations predicted by the fiducial model}, \(\sigma_{b_i}\) are the forecast errors derived from the Fisher matrix, and \(\Gamma(\zeta_i;v_g)\) are the theoretically expected correlations that are fully determined by $m_g$. 
Following the same procedure as in Section~\ref{sec:data-method}, we perform Bayesian parameter inference with \texttt{Cobaya} \cite{Torrado:2020dgo} using a uniform prior on \(m_g\) over the range {\(0 \leq \mg \leq h T_{\rm obs}^{-1}/c^2\)}. Convergence and burn-in are assessed following the same criteria described earlier.

The resulting posteriors for {$m_g$ are shown in the right panels of Fig.~\ref{fig:ska-forecast}. The $90\%$ credible upper limits below are listed as pairs (with \ac{CV}, without \ac{CV}):
\begin{align}
m_g^{90\%} &= (1.96\times10^{-24},\,2.24\times10^{-25})\,{\rm eV}/c^2 \quad (10\,{\rm yr})\,,\label{eq:ska-1}\\
m_g^{90\%} &= (8.00\times10^{-25},\,5.41\times10^{-26})\,{\rm eV}/c^2 \quad (20\,{\rm yr})\,,\label{eq:ska-2}\\
m_g^{90\%} &= (2.75\times10^{-25},\,9.27\times10^{-27})\,{\rm eV}/c^2 \quad (50\,{\rm yr})\,.\label{eq:ska-3}
\end{align}}
Thus, the constraints on $\mg$ are expected to be progressively tighter as the observational baseline increases, primarily due to the reduction in $\sigma_{b_i}$.
{For the fiducial forecasts including \ac{CV}, these projected limits improve on present \ac{PTA} bounds. The 10-year forecast is a few times tighter than the current \ac{NG} 15-year result, while the 50-year forecast improves it by more than one order of magnitude \cite{Wang:2023div,Wu:2023rib}. They also improve on the current \ac{CPTA} DR1 and \ac{LVK} bounds of order $10^{-23}\,{\rm eV}/c^2$ \cite{Wang:2023div,LIGOScientific:2019fpa,LIGOScientific:2021sio,LIGOScientific:2026fcf}. The substantially tighter noise-only limits illustrate the impact of neglecting \ac{CV}.}

To assess how our forecasts depend on the assumed pulsar sample quality, we consider an alternative, more conservative scenario. Instead of assigning $0.1\,\mu\mathrm{s}$ timing accuracy to all 174 pulsars, we adopt a two-population model, i.e., 50 ``golden'' pulsars with $\sigma_a = 0.1\,\mu\mathrm{s}$ and the remaining 124 pulsars with a degraded timing accuracy of $\sigma_a = 0.5\,\mu\mathrm{s}$. The sky positions are drawn isotropically for both populations, and all other noise parameters follow the same distributions as in the optimistic case. This configuration approximates a more realistic \ac{SKA}--\ac{PTA} pulsar sample, where only a subset of millisecond pulsars achieve the highest timing precision due to intrinsic flux-density limitations and scintillation effects \cite{Gitika:2025nqe}.

{Repeating the Fisher analysis under this conservative setup, the $90\%$ credible upper limits are listed as pairs (with \ac{CV}, without \ac{CV}):
\begin{align}
m_g^{90\%} &= (2.57\times10^{-24},\,4.63\times10^{-25})\,{\rm eV}/c^2 \quad (10\,{\rm yr})\,,\label{eq:mgpess1}\\
m_g^{90\%} &= (1.00\times10^{-24},\,1.15\times10^{-25})\,{\rm eV}/c^2 \quad (20\,{\rm yr})\,,\label{eq:mgpess2}\\
m_g^{90\%} &= (3.15\times10^{-25},\,2.16\times10^{-26})\,{\rm eV}/c^2 \quad (50\,{\rm yr})\,.\label{eq:mgpess3}
\end{align}}
{For the fiducial forecasts including \ac{CV}, these limits are approximately a factor of $1.1$--$1.3$ weaker than those obtained in the optimistic scenario,} indicating that while the pulsar sample quality does affect the absolute sensitivity, the overall improvement over current PTA constraints remains robust. Even in the conservative case, the 50-year \ac{SKA}--\ac{PTA} observation would reach $m_g \sim 3\times10^{-25}\,\mathrm{eV}/c^{2}$, still representing an order-of-magnitude improvement over the best current limits from NANOGrav \cite{Wang:2023div,Wu:2023rib}. This exercise demonstrates that the qualitative conclusion of this work, that \ac{SKA}--\ac{PTA} will substantially advance graviton-mass tests, is insensitive to the detailed assumptions about the pulsar timing precision distribution.

We further test the reference-frequency approximation used in the analysis by retaining the full frequency dependence of the massive-graviton response in the 10-year conservative \ac{SKA} configuration. In this controlled comparison, the same forecasting procedure is applied to both the single-frequency approximation, evaluated at $f_{\rm ref}=1/T_{\rm obs}$, and the frequency-dependent response evaluated at $f_k=k/T_{\rm obs}$ for $k=1,\ldots,30$. The resulting $90\%$ upper limits are $3.51\times10^{-24}$ and $9.16\times10^{-24}\,{\rm eV}/c^2$, respectively. Thus, retaining the full frequency dependence weakens the constraint by a factor of approximately $2.6$. The published \ac{MPTA} angular-correlation products do not provide the pair--frequency information required to perform the same frequency-dependent analysis for the current data, and the \ac{MPTA} limits therefore remain conditional on the $f_{\rm ref}=1/T_{\rm obs}$ approximation.

\section{Conclusions and discussion}
\label{sec:discussion-conclusion}

We have used the \ac{MPTA} 4.5-year angular-correlation measurements to constrain the graviton mass in the nanohertz \ac{GW} band. The analysis combines the massive-graviton dispersion relation with the corresponding modification of the tensor \ac{PTA} response, and applies the same Bayesian inference pipeline to the DATA, ER, and ALT noise configurations. This allows the mass constraint to be tested against the main noise-modelling ambiguity in the current \ac{MPTA} data set, rather than being quoted from a single assumed noise model.

The principal results are the 90\% credible upper limits on $\mg$ reported in Eqs.~\eqref{eq:mg_DATA_Q}--\eqref{eq:mg_ALT_Q}. The corresponding limits obtained after adding independent monopolar and dipolar components are given in Eqs.~\eqref{eq:mg_DATA_E}--\eqref{eq:mg_ALT_E}. The close agreement between these two sets of limits shows that the inferred graviton-mass bound is controlled by the quadrupolar angular structure, not by leakage into the additional monopole or dipole terms. Among the three \ac{MPTA} noise configurations, DATA gives the strongest limit. The weaker ER result follows from the anomalous red-noise component assigned to PSR\,J2129-5721, while ALT restores a constraint close to DATA after removing that component. The comparison therefore identifies the pulsar-noise treatment as the dominant practical limitation of the present \ac{MPTA} mass bound. {We note that the \ac{MPTA} angular-correlation measurements used here are obtained with the pulsar-noise parameters fixed at their posterior maxima. The uncertainties in these parameters are therefore not propagated in our analysis and may affect the inferred inter-pulsar correlation pattern.}

The \ac{SKA}--\ac{PTA} forecast demonstrates the expected progression from a present measurement to a precision nanohertz test of \ac{GR}. For 10-, 20-, and 50-year observing baselines, the projected 90\% upper limits {in the fiducial forecasts including \ac{CV}} reach the $10^{-24}$--$10^{-25}\,{\rm eV}/c^2$ range, {with the corresponding results with and without \ac{CV}} shown in Eqs.~(\ref{eq:ska-1})--(\ref{eq:ska-3}) (optimistic forecast) and Eqs.~(\ref{eq:mgpess1})--(\ref{eq:mgpess3}) (conservative forecast). The improvement is driven by both the smaller angular-correlation errors and the longer time baseline, which lowers the mass scale associated with a detectable departure from luminal propagation. The forecasts therefore show that \ac{SKA}--\ac{PTA} observations can move direct \ac{PTA} graviton-mass tests beyond the sensitivity of current \ac{PTA} data sets.

A direct quantitative comparison across different observational bands reveals the complementary nature of these constraints. For a massive graviton, the modification to the GW dispersion relation induces a frequency-dependent phase shift for ground-based interferometers. This makes LVK constraints primarily sensitive to the cumulative propagation effect over cosmological baselines, yielding $m_g \lesssim 10^{-23}\,\mathrm{eV}/c^{2}$ from binary coalescences \cite{LIGOScientific:2019fpa,LIGOScientific:2021sio,LIGOScientific:2026fcf}. 
Planned space-based detectors, e.g., the \ac{LISA}, are expected to improve the current graviton-mass constraints by one or two orders of magnitude \cite{Gao:2022hho,Luo:2025ewp}. 
{In contrast, the PTA constraints considered here arise from the deformation of the angular correlation pattern by massive-graviton dispersion.} Crucially, each band tests the same dispersion relation but with different source distances, frequency ranges, and systematic error budgets. {Null results across these bands remain consistent with a massless graviton, while any tension would require careful assessment of the corresponding response models and systematic uncertainties.}

We have also examined the impact of \ac{CURN} and stochastic signal variance, including \ac{CV}, on the inferred graviton-mass constraints. The present analysis assumes an isotropic stochastic background; possible anisotropies can modify the inter-pulsar angular-correlation pattern and may therefore affect the inferred constraints. A full treatment of anisotropic backgrounds is left for future work.

Within this broader program, the present \ac{MPTA} constraint is important because it is an independent nanohertz measurement based on a southern-sky \ac{PTA} data set and an angular-correlation observable \cite{Miles:2024seg}. It is competitive with the current \ac{NG} and \ac{CPTA} graviton-mass limits \cite{Wang:2023div}, and it tests the robustness of the inference under alternative treatments of pulsar noise. Continued \ac{MPTA} timing, joint analyses across \ac{PTA} collaborations, and the eventual \ac{SKA} pulsar sample will improve the angular coverage and reduce the statistical uncertainty of the correlation curve. These developments will make the nanohertz band a central part of multi-band tests of gravity, alongside ground-based and space-based interferometers.

In summary, the \ac{MPTA} 4.5-year data yield graviton-mass limits consistent with \ac{GR} and comparable to other current direct \ac{GW} constraints. However, unlike cosmological bounds which rely on the expansion history \cite{DeFelice:2023bwq}, our \ac{PTA} bounds are direct dynamical tests based on the propagation speed and angular correlation of \acp{GW}, thus providing an independent and complementary constraint that is less susceptible to the dark-energy model assumptions. The \ac{SKA}--\ac{PTA} forecast indicates that future nanohertz observations can improve the current \ac{PTA} limits by more than one order of magnitude, reaching $\mg$ sensitivities near $10^{-25}\,{\rm eV}/c^2$ including \ac{CV}. The combination of current \ac{MPTA} measurements and future \ac{SKA}--\ac{PTA} sensitivity therefore provides a clear route from evidence for nanohertz gravitational radiation to precision tests of the fundamental law of gravity.
\\
\bigskip

\begin{acknowledgements}
Z.C.Z. is supported by the National Key Research and Development Program of China (Grant No. 2021YFC2203001). 
S.W. is supported by the National Natural Science Foundation of China (Grant Nos. 12675065, 12533001, 12175243). 
This study was supported by the Advanced Computation Center of Hangzhou Normal University and the High-performance Computing Platform of China Agricultural University.
\end{acknowledgements}

\bibliographystyle{spphys}
\bibliography{references}

\end{document}